\documentstyle [12pt]{article}
\newcommand{\beq}{\begin{equation}}
\newcommand{\eeq}{\end{equation}}
\newcommand{\beqar}{\begin{eqnarray}}
\newcommand{\eeqar}{\end{eqnarray}}
\newcommand {\nono} {\nonumber \\} 

\newcommand {\bear} [1] {\begin {array} {#1}}
\newcommand {\ear} {\end {array}}

\newcommand {\eqr} [1] 
	{(eq.~\ref {eq:#1})}

\newcommand {\myI} [1] {\int \! #1 \,}

\newcommand {\wg} {\wedge}

\newcommand {\beqarn} {\begin{eqnarray*}}
\newcommand {\eeqarn} {\end{eqnarray*}}

\newcommand {\sepe} {\;\;\;\;\;\;\;\;}
\newcommand	{\come} {\;\;\;\;}

\newcommand	{\paren} [1] {{\left( #1 \right)}}
\newcommand	{\brak} [1] {{\left[ #1 \right]}}

\newcommand {\mrm} [1] {\mathrm {#1}}

\newcommand {\ch} {{\mbox {$\mrm{ch}$}}}

\newcommand {\myref} [1]	%
	{%
	\begin{thebibliography} {99}	%
			{#1}	%
	\end {thebibliography}}

\def\emph#1{{\em #1}}
\def\mathbf#1{{\bf #1}}
\def\mathrm#1{{\rm #1}}
\def\mathit#1{{\it #1}}

\newcommand {\half}	{\frac 1 2}

\catcode`\@=11
\@addtoreset{footnote}{section}
\catcode`\@=12

\newcounter	{exinsert}	[subsection]

\renewcommand {\theexinsert}	%
{	\thesubsection.\arabic{equation}}

\renewcommand {\theequation} {\thesection.\arabic{equation}}

\catcode`\@=11
\@addtoreset{equation}{section}
\catcode`\@=12

\newenvironment {exinsert} [1]	%
{	
	\begin {quotation}	%
	\refstepcounter {equation}	%
	{\bf {#1}} \theequation. \,\,	%
}	
{	
	\end {quotation}	
}

\newcommand {\bprop} 
{\begin {exinsert} {Proposition}
}
\newcommand {\eprop} {\end {exinsert} }

\newcommand {\bexe} {\begin {exinsert} {Exercise}}
\newcommand {\eexe} {\end {exinsert} }

\newcommand {\bexa} {\begin {exinsert} {Example}}
\newcommand {\eexa} {\end {exinsert} }

\newcommand {\nterm} [1] {{\emph {#1}}}

\newcommand {\literal} [1] {{ \mbox {$\mrm {#1}$}}}

\newcommand {\mod} [1] {{\;(\literal {mod} #1)}}

\newcommand {\Spin} {{\literal {Spin} }}

\newcommand {\gvary} {{\delta_g}}

\newcommand {\Kt} {{\literal {K3}}}

\newcommand {\lrfloor} [1] {{\lfloor {#1} \rfloor}}

\newcommand {\naiveq} {{\stackrel{naive}=}}

\newcommand {\ofe} [1] {{ \frac {#1} {48}}}
\newcommand {\oofe} {{ \ofe 1 }}

\newcommand {\nil} {\emptyset}
\newcommand {\swap} {\leftrightarrow}

\newcommand {\grad} {\nabla}

\def\jou#1,#2,#3,{{\sl #1\/ }{\bf #2} (19#3)\ }
\def\ibid#1,#2,{{\sl ibid.\/ }{\bf #1} (19#2)\ }
\def\am#1,#2,{{\sl Acta. Math.\/ } {\bf #1} (19#2)\ }
\def\annp#1,#2,{{\sl Ann.\ Physics\/ }{\bf #1} (19#2)\ }
\def\cmp#1,#2,{{\sl Comm.\ Math.\ Phys.\/ }{\bf #1} (19#2)\ }
\def\invm#1,#2,{{\sl Invent.\ Math.\/ }{\bf #1} (19#2)\ }
\def\cqg#1,#2,{{\sl Class.\ Quantum Grav.\/ }{\bf #1} (19#2)\ }
\def\ijmpa#1,#2,{{\sl Int.\ J.\ Mod.\ Phys.\/ }{\bf A#1} (19#2)\ }
\def\jhep#1,#2,{{\sl JHEP\/ }{\bf #1} (19#2)\ }
\def\jmp#1,#2,{{\sl J.\ Geom.\ Phys.\/ }{\bf #1} (19#2)\ }
\def\jams#1,#2,{{\sl J.\ Diff.\ Geom.\/ }{\bf #1} (19#2)\ }
\def\jdg#1,#2,{{\sl J.\ Amer.\ Math.\ Soc.\/ }{\bf #1} (19#2)\ }
\def\jpa#1,#2,{{\sl J.\ Phys.\ A.\/ }{\bf #1} (19#2)\ }
\def\grg#1,#2,{{\sl Gen.\ Rel.\ Grav.\/ }{\bf #1} (19#2)\ }
\def\mpla#1,#2,{{\sl Mod.\ Phys.\ Lett.\/ }{\bf A#1} (19#2)\ }
\def\nc#1,#2,{{\sl Nuovo Cim.\/ }{\bf #1} (19#2)\ }
\def\npb#1,#2,#3{{\sl Nucl.\ Phys.\/ }{\bf B#1} (19#2)\ }
\def\plb#1,#2,{{\sl Phys.\ Lett.\/ }{\bf #1B} (19#2)\ }
\def\pla#1,#2,{{\sl Phys.\ Lett.\/ }{\bf #1A} (19#2)\ }
\def\pnas#1,#2,{{\sl Proc.\ Nat.\ Acad.\ Sci.\/ }{\bf #1} (19#2)\ }
\def\prev#1,#2,{{\sl Phys.\ Rev.\/ }{\bf #1} (19#2)\ }
\def\prd#1,#2,{{\sl Phys.\ Rev.\/ }{\bf D#1} (19#2)\ }
\def\prl#1,#2,{{\sl Phys.\ Rev.\ Lett.\/ }{\bf #1} (19#2)\ }
\def\prpt#1,#2,{{\sl Phys.\ Rept.\/ }{\bf #1C} (19#2)\ }
\def\ptp#1,#2,{{\sl Prog.\ Theor.\ Phys.\/ }{\bf #1} (19#2)\ }
\def\ptpsup#1,#2,{{\sl Prog.\ Theor.\ Phys.\/ Suppl.\/ }{\bf #1} (19#2)\ }
\def\rmp#1,#2,{{\sl Rev.\ Mod.\ Phys.\/ }{\bf #1} (19#2)\ }
\def\sm#1,#2,{{\sl Selec. Math.\/ }{\bf #1} (19#2)\ }
\def\yadfiz#1,#2,#3[#4,#5]{{\sl Yad.\ Fiz.\/ }{\bf #1} (19#2) #3%
\ [{\sl Sov.\ J.\ Nucl.\ Phys.\/ }{\bf #4} (19#2) #5]}
\def\zh#1,#2,#3[#4,#5]{{\sl Zh.\ Exp.\ Theor.\ Fiz.\/ }{\bf #1} (19#2) #3%
\ [{\sl Sov.\ Phys.\ JETP\/ }{\bf #4} (19#2) #5]}

\def\atiyah	{M.~F.~Atiyah}
\def\alvarez	{O.~Alvarez}
\def\beckerk	{K.~Becker}
\def\beckerm	{M.~Becker}
\def\beckerd	{\beckerk, \beckerm}
\def\bershadsky	{M.~Bershadsky}
\def\callan	{C.~G.~Callan}
\def\douglas	{M.~R.~Douglas}
\def\ginsparg	{P.~Ginsparg}
\def\green	{M.~B.~Green}
\def\harvey	{J.~A.~Harvey}
\def\hori	{K.~Hori}
\def\agaume	{L.~Alvarez-Gaume}
\def\oz		{Y.~Oz}
\def\moore	{G.~Moore}
\def\morrison	{D.~R.~Morrison}
\def\ooguri	{H.~Ooguri}
\def\sadov	{V.~Sadov}
\def\sen	{A. Sen}
\def\polchinski	{J.~Polchinski}
\def\schwarz	{J.~H.~Schwarz}
\def\singer	{I.~M.~Singer}
\def\strominger	{A.~Strominger}
\def\vafa	{C.~Vafa}
\def\wess	{J.~Wess}
\def\witten	{E.~Witten}
\def\zumino	{B.~Zuimino}
\def\zweibach	{B.~Zwiebach}

\def\zy		{Z.~Yin}

\newcommand {\bsp} {\;\:}

\newcommand {\hAud} [2] {{\frac {\hat A[{#1}]} %
			{\hat A[{#2}]} }}

\newcommand {\hAtnd} [1] {{\hAud {T(M_{#1})} {N(M_{#1})}}}

\newcommand {\hAtni} {{\hAud {T(M_1) \cap T(M_2)} %
		{ N(M_1) \cap N(M_2)}}}

\newcommand {\Qind} {{ Q_{\literal{ind}}}}

\begin {document}

\begin{titlepage}

\begin{center}
hep-th/9710206 \hfill PUPT-1732 \\
\hfill LBNL-40347, UCB-PTH-97/25        
\vskip 1 cm
{\LARGE \bf  Anomalies, Branes, and Currents
\\}
\vskip 1 cm 
{Yeuk-Kwan E. Cheung$^1$\footnote 
	{email:  cheung@viper.princeton.edu}
and Zheng Yin$^2$\footnote {email:  zyin@thsrv.lbl.gov}}\\
\vskip 0.75cm
{\sl 	$^1$Department of Physics \\
	Princeton University \\
	Princeton, NJ  08544-0708 \\
	\vskip 0.5cm
	$^2$Department of Physics \\
	University of California \\
	Berkeley, CA 94720-7300 \\
	{\it and} \\
	Mail Stop 50A-5101 \\
	Lawrence Berkeley National Laboratory \\
	1 Cyclotron Road, \\
	Berkeley, CA  94720 \\
}
\end{center}
\vskip 0.5 cm
\begin{abstract}
When a D-brane wraps around a cycle of a curved manifold,
the twisting of its normal bundle can induce chiral asymmetry
in its worldvolume theory.
We obtain the general form of the resulting anomalies for
D-branes and their intersections.
They are not cancelled among themselves, and 
the standard inflow mechanism does not apply at 
first sight because of their apparent lack of 
factorizability and the apparent 
vanishing of the corresponding 
inflow.  We show however after taking into 
consideration the effects of the nontrivial topology of 
the normal bundles, the anomalies can be transformed into 
factorized forms and precisely cancelled by finite 
inflow from the Chern-Simons actions for the D-branes 
as long as the latter are well defined.  
We then consider examples in type II compactifications 
where the twisting of the normal bundles occurs 
and calculate the changes in the
induced Ramond-Ramond charges on the D-branes.
\end{abstract}

\end{titlepage}

\section 	{Introduction}

	In recent studies of string theory, 
brane configurations play a very important role.  The low
energy physics of such configurations are that of 
field theories, which often  possess both 
gauge and global symmetries.  In such constructions, 
some global symmetries, usually the R symmetries  
that act on the supercharges,
originate from the rotation symmetry
of the bulk string theory restricted to the normal bundles of
the branes.  They are gauged in the bulk spacetime and 
therefore must be free of anomalies, just 
as the symmetries gauged on the branes.  
However, there is generically chiral
asymmetry with respect to these global symmetries 
on a D-brane or the intersection of a pair 
of D-branes, known as an \nterm {I-brane}.  It brings about 
pure and mixed anomalies involving these global symmetries 
in the effective brane worldvolume theory.
If this were the only story, such brane 
configurations would be inconsistent.

	The mechanism to cancel the anomaly in
an otherwise anomalous theory is to compensate it with an  
``anomalous''  
variation of the classical action.  An 
example is the Green-Schwarz mechanism for type I and 
heterotic string theories \cite {gs}.  More 
generally, the anomalous theory can be embedded in a higher 
dimensional theory.  The anomalous 
variation of the classical action of the bigger 
theory is localized at (``flows'' to) the worldvolume for 
the anomalous theory and cancels 
its anomaly, hence the name \nterm {anomaly inflow} 
\cite{ch,bh}.
More recently it has been applied to derive the 
Chern-Simons type of
actions on D-branes, whose classical variations cancel 
the Yang-Mills and gravitational anomalies that appear on a 
certain class of I-branes \cite {ghm}.  However, 
there are additional anomalies associated with the 
global R symmetries as mentioned earlier.  They exist for 
generic D-branes and their intersections.
If D-branes are
wrapped around nontrivial cycles of a curved 
compactification
manifold \cite {bsv2, ooy, bbmooy}, 
the anomalies can manifest themselves as 
nonvanishing variation of the effective action 
under a local gauge transformation.  Such scenarios 
have appeared in studies of string dualities \cite{bsv1, ov1} as
well as field theory dualities \cite
{bv, ov2, vz, ho}.  They have also found use in studying 
topological field theories \cite {bsv2, top1, top2}.  However, 
anomaly cancellation for them 
has not been investigated until now.

In generalizing the inflow method to such
cases, one inevitably runs up against a serious obstruction.  
Factorizability of an anomaly, as
defined precisely later, is crucial for it to be cancelled 
via the inflow 
mechanism.  However, for the additional anomalies we 
study, factorizability is apparently
lost.  To recover it we shall encounter a classic result
from differential topology\footnote
	{This result, the relation between Thom class and 
	Euler class, has also been used in a different context:
	anomaly analysis for the NS5-branes in type IIA 
	string theory and the 5-branes in M theory
	\cite{witea, bcr}.}.
It allows us to cancel the new anomalies in 
all cases as long as the D-brane Chern-Simons actions 
are well-defined.
 
The D-brane Chern-Simons actions derived in \cite {ghm} 
imply that
topological defects on D-branes carry their own Ramond-Ramond 
charges 
determined by their topological 
(``instanton'') numbers. 
This observation has far reaching consequences \cite {Tsi, Tbwb,
Tgfab}.  To cancel the new anomalies that we study, 
the Chern-Simons actions are modified. 
This can change the induced 
Ramond-Ramond charges on a D-brane if it is 
wrapped around some cycle of
a nontrivial compactification manifold.

	The plan of this paper is as follows.  In section 2 
we discuss how the inflow mechanism works.  In addition 
to a review of some known results, we shall uncover 
subtleties in the choice of the 
kinetic action 
for the Ramond-Ramond field that have not been addressed 
in the literature.
We also define carefully the notion of brane current.  
For describing flat D-branes, it is just a very convenient 
notation, but in the anomaly 
cancellation considered later in this paper, 
it plays an essential 
role.  In section 3 we consider the chiral asymmetry induced 
by twisting the normal bundle and compute the resulting anomaly. 
We then point out the apparent obstruction to cancelling such 
anomaly.  
In section 4  this difficulty is overcome 
with the help of some interesting topological
information encoded in the brane current.  
Then in section 5 we give examples 
where the normal bundles of D-branes are nontrivial and 
calculate the induced Ramond-Ramond charge.  
In the appendix we 
comment on the relevance of brane stability and supersymmetry 
to our anomaly analysis.

\section {The Inflow Mechanism}

The inflow mechanism was originally discovered in the context of 
gauge theory \cite {ch}, where the action in spacetime 
has a gauge noninvariant term.  Its variation is concentrated 
on topological defects and cancels the anomalies 
produced by their chiral fermion 
zero modes.  It was recognized 
in \cite {ghm} that this mechanism also applies to the 
Yang-Mills and gravitational anomalies 
that arise for a certain class of intersecting D-branes 
in string theory.  
In this section, we present systematically the details of the 
inflow mechanism.  Although much of it is a review of the 
earlier results cited above, there are some salient departures.  
The most important one being our use of a kinetic action 
manifestly symmetric 
with respect to all 
Ramond-Ramond potentials.  
Its use is really required by the way the inflow mechanism 
works for D-branes and 
turns out to be 
important for reproducing the correct Ramond-Ramond charge.  

	As it shall become clear, an anomaly must be 
factorizable in an appropriate sense in order to be 
cancelled by inflow.  
One of the difficulty associated with the anomalies we consider 
in this paper is their apparent lack of factorizability, and 
the key to cancelling them involves rewriting 
them in a factorized form.

\subsection {Branes and Currents}

	Before discussing the detail of the inflow mechanism, 
we first introduce a notion that is very convenient here 
and will prove essential later.  
Usually a brane is introduced into the bulk theory
by adding to the bulk action \[	\int_M {\cal L}_M	\]
where $M$ is the $m$-dim worldvolume of the brane and
${\cal L}_M$ the Lagrangian density governing the dynamics 
on the brane.  One may rewrite this into an integration over
total (bulk) spacetime $X$, with the help of a
``differential form'' $\tau_M$, defined by 
\beq	\label {eq:def-tau-M}
	\int_M \zeta \equiv \int_X \tau_M \wg \zeta
\eeq
for all rank $m$ form $\zeta$ defined on $M$\footnote 
	{This definition makes sense because any form 
	$\zeta$ on $M$ can be extended to be a form on
	$X$ by a suitable bump function with support 
	on a tubular neighborhood of $M$.  Conversely, 
	if $\zeta$ is a form defined on $X$ to start with, 
	pull-back to $M$ is implicit on the LHS of 
	\eqr {def-tau-M}, as in
	similar expressions throughout this paper.}.  
Thus 
the rank of $\tau_M$ is equal to the codimension of $M$ 
in $X$.  To be precise, \eqr {def-tau-M} defines $\tau_M$ as an
element in the dual of the space of forms, known to 
mathematicians as the space of \nterm {currents}\,\cite {gh3}.
Currents are differential-form analogue of distributions;
likewise, $\tau_M$ is the generalization of Dirac's delta
function\footnote
	{In this language, a delta function in $R^d$
	is really a rank $d$ current that maps a function
	(0-form) into a number.}.
Obviously, $\tau_M$ must have singular support on $M$ and
integrate to $1$ in the transverse space of $M$.

	In \eqr{def-tau-M}, the form $\zeta$ is allowed
to be any form on $M$.  If instead it is restricted to be
closed, the same equation only defines a cohomology class
$[\tau_M]$, known as the \nterm {Poincare dual} of $M$.  It
contains topological information about $M$.  
$\tau_M$ can be defined as a particular representative of
$[\tau_M]$ that is supported only on $M$.

	In this paper, we shall call $\tau_M$ the brane current 
associated with a brane wrapped around $M$, for a very
physical reason.  
For illustration, consider 
a $d$-dim gauge theory with a conventional 2-form
field strength $F$. Let $M$ be the worldline trajectory of an
electrically charged particle embedded in the total
spacetime $X$.  The kinetic term for the gauge field $F$ is 
\beq
	S_{\literal{gauge}} = - \half \int_X F \wg *F.
\eeq
The coupling of the potential to the electron is 
\beqar	\label {eq:worldline}
	S_{\literal {matter}} & = & - \int_M A \nono
		& = & - \int_X \tau_M \wg A.
\eeqar
Then the equation of motion for $A$ yields 
\beqar	\label {eq:j-brane-current}
	* j_{ele} & \equiv &  d*F \nono 
	 & = & (-1)^d \tau_M.
\eeqar
So the usual physical current (source) is related to $\tau_M$ 
by a Hodge $*$
operation.  Similarly, if $\hat M$ is the $(d-3)$-dim
worldvolume of a magnetically charged object, the
Bianchi identity would read something like 
\[	* j_{mag} \equiv d F = \pm \tau_{\hat M}.	\]
Now return to string theory.  Let M be the worldvolume of 
a D-brane.
It couples to the Ramond-Ramond potential $C$ of the 
appropriate rank just as in \eqr {worldline} but with $A$ 
replaced by $C$.  Then \eqr {j-brane-current} gives 
the definition of the brane current $\tau_M$ with $F$ replaced 
by the appropriate Ramond-Ramond field strength $H$.

On $M$, the tangent bundle $T(X)$ of the total spacetime 
$X$, decomposes into the Whitney sum of $T(M)$ and
$N(M)$, the tangent and normal bundles to $M$ respectively. 
Note that within each fiber of $N(M)$ \eqr {j-brane-current} 
is just the usual Poisson equation.  
Its RHS has
Dirac's $\delta$-type singular support on the zero section.  
Thus $\tau_M$ can be constructed \emph {locally} as 
\beq	\label {eq:naive-current}
	\tau_M \naiveq \delta (x^1) dx^1 \wg \cdots 
	\wg \delta (x^{\dim N(M)}) d x^{\dim N(M)},
\eeq
where $x^\mu$ are Gaussian normal coordinates in the
transverse space of $M$, or equivalently 
Cartesian coordinates in the 
fiber of $N(M)$.  We emphasize that this expression
is naive and in general ill-defined globally.

	Now consider the intersection $M_{1 2} \equiv M_1
\cap M_2$ of two brane-worldvolumes $M_1$ and $M_2$.  In the 
literature $M_{1 2}$ has been called I-brane.  For simplicity 
we shall 
concentrate on I-branes from intersections at right angle, 
but the results apply to other cases as well\footnote 
	{The basic reason is that the relevant 
	quantum numbers of the massless fermions 
	are determined by 
	$T(M_1) \cap T(M_2)$  and $N(M_1) \cap N(M_2)$, 
	which are well defined even for oblique intersections.}. 
The right angle condition implies that on the
I-brane $M_1 \cap M_2$, the tangent bundle of the total 
spacetime $T(X)$ decomposes as follows:
\beqar	\label {eq:TX-factorize}
	T(X) & = & 
	  T(M_1) \cap T(M_2) 
	  \oplus T(M_1) \cap N(M_2) \nono
	& & \oplus N(M_1) \cap T(M_2) 
	\oplus N(M_1) \cap N(M_2),
\eeqar
where $\cap$ denotes fiberwise set theoretic intersection.
It is clear that
\beq
	T(M_{1 2}) = T(M_1) \cap T(M_2)
\eeq 
and
\beq
	N(M_{1 2}) = T(M_1) \cap N(M_2) 
	\oplus N(M_1) \cap T(M_2) 
	\oplus N(M_1) \cap N(M_2).
\eeq
Then\,\eqr {naive-current} implies that 
\beqar	\label {eq:naive-intersection}
	\tau_{M_1} \wg \tau_{M_2} &=& \tau_{M_{1 2}} 
	  \sepe \mbox {\hfill if $N(M_1) \cap N(M_2) = \O$,} \nono
	&{\naiveq}& 0 \sepe \mbox {\hfill otherwise,}
\eeqar
where in the second line we have used the anticommutivity of
exterior multiplication.  Here again we emphasize that the second
equation is naive, because it uses the naive expression
\eqr {naive-current}.  The correct statement and its
important implication will be given in section 4.
Intersections on which $N(M_1) \cap N(M_2) = \O$  are known as 
\emph {transversal}.  

\subsection {The Inflow}

	Suppose the anomaly on an I-brane $M_{1 2}$ 
can be written in the following form:
\beq	\label {eq:model-anomaly}
	I_{1 2} = \pi \int \tau_{M_{1 2}} \wg 
	\left(Y_1 \wg \tilde Y_2 + Y_2 \wg \tilde 
     	Y_1\right)^{(1)},
\eeq
where $Y_i$ and $\tilde Y_i$, $i = 1, 2$, are some invariant
polynomials of the Yang-Mills field strengths and 
gravitational curvatures defined on $M_i$.  The expression  
${Z}^{(1)}$ denote the Wess-Zumino descent \cite {wz, zlec} 
of an invariant curvature 
polynomial $Z$: if $N$ is the constant part of $Z$,
\[Z \equiv  N + Z_{0},\]
and $Z^{(0)}$ is its secondary characteristic, 
\[	Z_{0} \equiv d Z^{(0)},	\]
then the gauge variation of $Z^{(0)}$ is 
\[	\gvary Z^{(0)} \equiv d Z^{(1)}.	\]
$Y_i$ and $\tilde Y_i$ must be defined entirely 
by the D-branes wrapping $M_i$.  For example, $Y_1$'s 
dependence on the gravitational 
curvature from $T(M_1)$ and $N(M_1)$ may be different, but 
it must not distinguish, say, between the contributions from 
$T(M_1) \cap T(M_2)$ and $T(M_1) \cap N(M_2)$.
In this paper such an anomaly is called factorizable.  
	
	To cancel the anomaly \eqr{model-anomaly}, one 
introduces the
following ansatz for a Chern-Simons type action on D-branes 
\cite {ghm}\footnote
	{T-duality relates the charge $\mu$ for D-branes of 
	different dimensions.  With a suitable choice 
	for the unit of length, they are all equal \cite {pol}.}:
\beq	\label {eq:c-s-action}
	- \frac {\mu} 2 
	\sum_i \int_{M_i} N_i C- (-1)^q H \wg Y_i^{(0)}
	= - \frac {\mu} 2 \sum_i \int_X \tau_{M_i} 
	\wg (N_i C - (-1)^q H \wg Y_i^{(0)}).
\eeq
Here $q$ is 1 for IIA and 0 for IIB string theory.
$i$ labels the D-brane wrapping worldvolume $M_i$, whose
brane current is $\tau_{M_i}$.  $N_i$ is the constant 
part of $Y_i$.  Anomaly computation in section 3 will show 
that it is the multiplicity of the D-branes wrapping $M_i$.
$\mu$, rather than $\frac {\mu} 2$ as one would naively 
expect, is the brane charge, for reason to be explained shortly.  
$C$ and $H$ are the 
formal sums of all the 
Ramond-Ramond antisymmetric tensor potentials and field strengths
respectively.  Integration automatically picks out products 
of forms with the appropriate total rank.  
In the following we shall often denote 
by $Z_{(n)}$ the rank $n$ part of any formal sum $Z$.  For 
example, 
\[
	C = C_{(1)} + C_{(3)} + C_{(5)} + C_{(7)} + C_{(9)}
\]
for type IIA string theory and 
\[
	C = C_{(0)} + C_{(2)} + C_{(4)} + C_{(6)} + C_{(8)}
\]
for type IIB string theory.	
It is important to remark that unlike the usual Chern-Simons 
action, in \eqr {c-s-action} one cannot use integration 
by parts to reduce the RHS to the more uniform expression 
of $- \frac {\mu} 2 \int_{M_i} C \wg Y_i$.  The
reason is, as we shall see, $d H_{(n)} \neq 0$, even 
away from any magnetic D(8-n) brane.  
So $H$ has corrections to its usual
expression of $d C$: 
\beq	\label {eq:def-H-rough}
	H = d C + \cdots .
\eeq
Therefore a brane Lagrangian in the form of 
$- \frac \mu 2 C \wg Y$ is
different from \eqr{c-s-action} by some additional terms.  
In fact, only \eqr {c-s-action} can cancel the factorized 
anomaly \eqr {model-anomaly}.

	In a theory that treats electric and magnetic
potentials on equal footing, 
there could be ambiguity in deriving 
the equations of motion using the conventional 
kinetic action.  Since \eqr {c-s-action} explicitly 
involves both electric and magnetic sources, it must be 
understood to be part of an action that is a manifestly 
\emph {electro-magnetically symmetric}.  The detail of the action and 
its ramifications are interesting in their own rights
and presented 
in the next subsection.  The relevant 
results can be summarized as 
follows: given the coupling 
in \eqr {c-s-action}, with the factor of $\half$, 
the equations of motion are 
\beq	\label {eq:eom}
	d * H = \mu 
		\sum_i \tau_i \wg Y_i,
\eeq
and the Bianchi identities are 
\beq	\label {eq:b-id}
	dH = - \mu 
		\sum_i \tau_i \wg \tilde Y_i,
\eeq
with 
\beq	\label {eq:def-tilde-Y}
	\tilde Y_{j (l)} = 
		- (-1)^{\frac {dim(M_j) - q} 2} 
		(-1)^{l/2} {Y_j}_{(l)},
\eeq
\emph {without} any factor of $\half$!
Note $Y$ and $\tilde Y$ 
are in general different.  It will become apparent later 
that the factor $(-1)^{l/2}$ relates $\tilde Y$ 
to $Y$ by complex conjugation of the group representation 
of the associated Yang-Mills gauge group, 
while the factor $(-1)^{\frac {dim(M_j) - q} 2}$ chooses 
an orientation for the I-brane.

	The Bianchi identities \eqr {b-id} impose very strong 
conditions 
on the terms represented by $\cdots$ in \eqr{def-H-rough}.  
The minimal expression for $H$ is 
\beq	\label {def-H}
	H = dC - \mu (-1)^q \tau_{M_j} \wg \tilde Y_j^{(0)},
\eeq
where $\tilde N_j$ is the constant part of $\tilde Y_j$, and 
$\tilde Y_j^{(0)}$ its secondary characteristic 
(similar notations apply to the untilded $Y$'s).
Since the field strengths $H$ are physical observables, 
they must be
invariant under gauge 
transformations.  Thus $C$ must have compensating 
gauge variations:
\beq 
	\gvary C = 
	\mu \sum_j \tau_{M_j} 
		\wg \tilde Y_j^{(1)},
\eeq
where $\tilde Y_j^{(1)}$ is the Wess-Zumino descent 
of $\tilde Y_j$.

	Now we can compute the variation of 
\eqr {c-s-action} under gauge transformations to be 
\beqar	\label {eq:gvary-s}
\gvary  S & = & - \frac {\mu^2} 2 \sum_{i j} \int_X \tau_{M_i}\wg 
	\tau_{M_j} \wg \paren { \tilde Y_j^{(1)} N_i +
	\tilde Y_j \paren {Y_i}^{(1)}}	\nono
	&=& - \frac {\mu^2} 2 \sum_{i j} \int_X \tau_{M_i} \wg 
	\tau_{M_j} \wg \paren {Y_i \wg \tilde Y_j}^{(1)}.
\eeqar
For a particular pair of \emph {distinct} D-brane worldvolume 
$M_1$ and $M_2$, this gives an anomalous variation 
\beqar	\label	{eq:gvary-s-12}
	\gvary S_{1 2} & = & -  \frac {\mu^2} 2
		 \int_X \tau_{M_1}
	\wg \tau_{M_2} \paren { \tilde Y_2^{(1)} N_1  + 
	\tilde Y_2 \paren {Y_1}^{(1)}+\tilde Y_1^{(1)} N_2 +
	\tilde Y_1 \paren {Y_2}^{(1)}}\nono
	& = & - \frac {\mu^2} 2 
		\int_X \tau_{M_1} \wg \tau_{M_2} 
	\paren {Y_1 \wg \tilde Y_2 + Y_2 \wg \tilde 
     	Y_1}^{(1)}.
\eeqar
According to the first equation in \eqr {naive-intersection}, 
when $N(M_1) \cap N(M_2) = \emptyset$, this \emph {inflow} 
precisely cancels the anomaly \eqr {model-anomaly} if 
\beq	\label {eq:controversial}
	\frac {\mu^2} 2  = \pi.
\eeq
So the anomaly and inflow analysis also 
constitutes an independent verification 
of brane charge 
computed in \cite {pol-RR}.  The  factor of 
$\half$ in \eqr {c-s-action} relative to \eqr {eom} and 
\eqr {b-id} is crucial for agreement\footnote 
	{In \cite {ghm}, there was no factor of $\half$ 
	in the Chern-Simons action, but the total anomaly 
	was computed to be twice as large, so the same value for 
	$\mu$ was obtained.  We would like to 
	thank the authors of \cite {ghm} for useful 
	communications regarding this issue.}.

	The cases with $N(M_1) \cap N(M_2) = \emptyset$ were 
considered in \cite {ghm}.  When this does not hold, 
the second equation in \eqr {naive-intersection}  suggests 
that  
the inflow \eqr {gvary-s-12} vanishes.  However, we shall 
show in section 3 that on the corresponding I-branes there 
still exist anomalies.  Fortunately, in section 4, we shall 
find the correction to 
\eqr {naive-intersection} that keeps 
the inflow finite and cancels the anomaly.

\subsection {Electro-magnetically Symmetric Action}

	In this subsection we derive the equations of 
motion \eqr {eom} and justify the 
relative factor of $\half$ in \eqr {c-s-action}\footnote 
	{A similar factor of $\half$ in the coupling to 
	sources has also been suggested 
	recently in \cite {dght}.  However, the detailed 
	form of the 
	action used there seems 
	to be different.}.  
As mentioned earlier, this factor is essential for obtaining 
the correct brane charges required by string duality 
\cite {pol-RR, pol}.
The kinetic action for antisymmetric tensors 
we shall use is the one proposed in 
\cite {ss} for source free situations.  After 
we couple it to sources, it is well 
suited for \eqr {c-s-action} because it treats both 
electric and magnetic potentials on the same footing.  
The price to pay is the loss of manifest 
Lorentz invariance --- the action has only manifest 
rotation invariance in the spatial dimensions, although 
it possesses additional symmetries 
that reduce on shell to the usual Lorentz transformations
\cite {ss}.  More recently, there has been progress in 
covariantizing it\footnote
	{See, for example, \cite {sorokin}.}.  However, 
for the present discussion the simpler 
noncovariant version suffices.

	First consider just one electro-magnetic dual pair 
of RR fields $H_{(n)}$ and $\breve H_{(d-n)}$, where the 
subscripts, often omitted,  
denote the ranks of forms.  Their respective 
potentials are $C_{(n-1)}$ and $\breve C_{(d-n-1)}$.  
Now let
\beq
	C = \Phi + A
\eeq
and 
\beq
	H = E + B
\eeq
so that the components of $\Phi$ and $E$ consist of those of 
$C$ and $H$ respectively with a temporal index, while $A$ and 
$B$ have only spatial indices.  Similarly we can also 
decompose 
the spacetime exterior derivative $d$ into the 
spatial exterior 
derivative $\grad$ and the temporal part $d_t$:
\beq
	d = d_t + \grad,
\eeq
with 
\beq	\label {eq:nilpotence}
	\{d_t, \grad\} = 0 = d_t^2 = \grad ^2.
\eeq
Then 
\beqar
	E &=& d_t A + \grad \Phi. \\
	B &=& \grad A.
\eeqar
The analogy with the usual non-manifestly Lorentz 
covariant formulation of electrodynamics should be 
clear.  The same can be carried out for the dual fields: 
\beqar
	\breve H &=& \breve E + \breve B, \nono
	\breve C &=& \breve \Phi + \breve A.
\eeqar 

	Consider now the action \cite {ss}:
\beq	\label {eq:ss-action}
	S_{BE} = -\half \int (B \wg \breve E - E \wg \breve B 
		+ B \wg * B + \breve B \wg * \breve B).
\eeq
In the absence of sources, the fields satisfy the 
following Bianchi identities in light of \eqr {nilpotence}:
\beqar
	\label {eq:b-id-0-B}
	\grad B \bsp = &0& = \bsp \grad \breve B, \\
	\label {eq:b-id-0-E}
	d_t B + \grad E \bsp = &0& = \bsp 
		d_t \breve B + \grad \breve E.
\eeqar
By using the first of them one finds that the 
equations of motion for $\Phi$ 
and $\breve \Phi$ are trivially satisfied --- they only 
enter the action as parts of total exterior derivatives.  
This implies a larger set of gauge transformations than 
in the usual formulation:
\beqar	
	\label {eq:ss-gvary-A}
	\gvary A = \grad \Gamma, && 
		\gvary \breve A = \grad \breve \Gamma;\\
	\label {eq:ss-gvary-Phi}
	\gvary \Phi = \Psi, && 
		\gvary \breve \Phi = \breve \Psi
\eeqar
with independent $\Gamma$, $\breve \Gamma$, $\Psi$, 
and $\breve \Psi$.  The gauge transformations
\eqr {ss-gvary-Phi} allow $\Phi$ and $\breve \Phi$ 
to be set to 0, corresponding to the usual temporal gauge.
Applying \eqr {b-id-0-E}, the equations of motion 
for $A$ and $\breve A$ are found to be  	
\beqar
	\grad (\breve E + * B) &=& 0; \\
	\grad (E - (-1)^{n(d-n)} * \breve B) &=& 0
\eeqar
respectively: the expressions inside the parenthesis are 
closed.  By using the gauge transformations 
\eqr {ss-gvary-A}, one can choose a gauge 
so that they vanish:
\beqar	\label {eq:duality}
	\breve E &=& - * B; \nono
	E &=& (-1)^{n(d-n)} * \breve B.
\eeqar
They then give the duality relation between 
$H$ and $\breve H$.  Substituting them for the Bianchi 
identities \eqr {b-id-0-B} and \eqr {b-id-0-E} one 
finally recovers the 
conventional equations of motion for antisymmetric 
tensors:
\beqar	\label {eq:eom-0}
	\grad * E \bsp = &0& = 
		\bsp d_t * E + \grad * B; \nono
	\grad * \breve E  \bsp = &0& = 
		\bsp d_t * \breve E + \grad * \breve B.
\eeqar
Note that although the action \eqr {ss-action} is not 
Lorentz invariant, the equations of motion obtained from 
it are.  Furthermore, one can recover from 
\eqr {ss-action} the conventional 
action for one of the gauge potential, say $A$, 
in temporal gauge 
by solving the duality equation \eqr {duality} for its 
dual $\breve A$ and make 
the gauge choice 
\beq
	\Phi = 0 = \breve \Phi.
\eeq

	Now let us put in the sources.  In the conventional 
action formalism, where only one potential is used, the 
potential remains single valued 
if just electric sources are present.  When 
there is also magnetic source, the potential can only be 
defined over patches --- it is a connection of a nontrivial 
bundle \cite {wy}.  The Bianchi identities must be modified.  
When one switches to the dual description, the meaning of 
electric and magnetic sources are interchanged, as are 
the equations of motion and the Bianchi identities.  
In the symmetric formalism we use here, because both of the 
dual pair of potentials are used, some Bianchi identities 
must be modified whichever type of sources is introduced --- 
there is no longer a meaningful 
distinction between ``electric'' and 
``magnetic'' sources.  However they are called, 
the same set of equations for the field strengths 
must obtain  
in all three approaches if they are equivalent.

	Let the brane current for the the sources 
be proportional to  
\beqar
	\lambda = \omega + \sigma,\\
	\breve \lambda = \breve \omega + \breve \sigma,
\eeqar
with the decomposition into the temporal parts ($\omega$ and 
$\breve \omega$) and the 
spatial parts ($\sigma$ and $\breve \sigma$) 
understood.  They are normalized 
so that the Bianchi identities are now 
\beqar	\label {eq:b-id-1}
	\grad B &=& \breve \sigma,\nono
	d_t B + \grad E &=& \breve \omega; \nono
	\grad \breve B &=& \sigma,\nono
	d_t \breve B + \grad \breve E &=& \omega.
\eeqar

These brane currents also make a contribution, denoted 
by $S_j$, to the total action.  
One can derive the form of $S_j$
by using the modified Bianchi identities \eqr {b-id-1}.  
The equations of motion for $\Phi$ and 
$\breve \Phi$ require the dependence of $S_j$ on them to 
be
\beq
	S_j = \half \int \left( (-1)^{n+1} \Phi \wg \sigma
		+ \half (-1)^{(n+1)(d-n)} 
			\int \breve \Phi \wg \breve \sigma
		+ \cdots \right).
\eeq
This is necessary for the consistency of the theory and 
ensures that the gauge transformations \eqr {ss-gvary-Phi} 
continue to hold.  Note the factor of $\half$.  It comes 
from the same factor in \eqr {ss-action}.

	Turning now to the equations of motion 
for $A$ and $\breve A$, 
we demand that the duality relation \eqr {duality} holds 
again.  This completely fixes the dependence of 
$S_j$ on them:
\beq	\label {eq:sj}
	S_j = \half \int \left( (-1)^{n+1} A \wg \omega
		+ \half (-1)^{(n+1)(d-n)} 
			\int \breve A \wg \breve \omega
		+ \cdots \right).
\eeq
Now $S_j$ is completely determined and has 
a Lorentz invariant expression:
\beq
	S_j = \half \int \left( (-1)^{n+1} C \wg \lambda
		+ (-1)^{(n+1)(d-n)} \breve C \wg \breve \lambda
		\right).
\eeq
The conventional equations of motion are again determined 
from the Bianchi identities \eqr {b-id-1} and the duality 
relation \eqr {duality}:
\beqar	\label {eq:eom-1}
	\grad * \breve E &=& (-1)^{n(d-n)} 
		\breve \sigma,\nono
	d_t * \breve E + \grad *\breve B &=& 
		(-1)^{n(d-n)} \breve \omega; \nono
	\grad *E &=& - \sigma,\nono
	d_t *E  + \grad *B &=& - \omega.
\eeqar
When, say, $\breve \lambda = 0$, 
one can recover the 
conventional action in temporal gauge for $C$ just as for 
the source free case.  The resulting 
source term is found to be conventionally normalized, i.e. 
\emph {without} the factor $\half$.  When both an 
electric brane of charge $q_e$ and a magnetic brane 
of charge $q_m$ are present, deforming the worldvolume of, 
say, the electric brane around the magnetic 
brane by a complete revolution shifts the action 
\eqr {sj} by a constant.  The electric and magnetic parts of 
\eqr {sj} each makes an equal contribution of $\half q_e q_m$.  
Requiring $\exp (i S_j)$ to be single-valued reproduces 
the standard Dirac quantization: $q_e q_m = 2m\pi$.

	Finally, we shall write down the electro-magnetically 
symmetric action for the Ramond-Ramond fields, which is 
directly relevant for the inflow mechanism.  
In string theory, a Ramond-Ramond field strength $H_{(n)}$ and
its dual $*H_{(n)}$ appear on equal footing.  
The formal sum $H$ actually includes all electro-magnetic dual 
pairs of Ramond-Ramond field strengths, and so does $*H$.  
To find their
relation, recall that
these field strengths can be defined as follows in terms of 
the decomposition of 
bispinors:
\beq 	H_{\mu_1 \ldots \mu_n} 
	= S_L^T \Gamma_{\mu_1} \ldots \Gamma_{\mu_n} S_R.
\eeq
Here $S_L$ has positive $Spin(1,9)$ chirality, while $S_R$ 
has positive or negative chirality for IIB and IIA string
respectively.  It is straightforward to infer from this 
\beq	H_{(n)} = (-1)^{(n+q-1)/2}*(H_{(10-n)}).
\eeq
Recall that $q$ is 0 for IIB and 1 for IIA theory.  
These duality relations can be obtained from the action 
\beq	\label {eq:symmetric-II-action}
	S_{BE} = -\half \myI {d^{10}x} 
		\sum_n \left( (-1)^{(n-q+1)/2} 
			B_{(n)} \wg E_{(10-n)}
			+ B_{(n)} \wg * B_{(n)} \right).
\eeq
Then if $S_j$ is the Chern-Simons coupling in \eqr {c-s-action}, 
it can be shown that the Bianchi identities must be 
\eqr {b-id} and the equations of motion must be \eqr {eom}.

\section {Brane Anomalies}

As usual, the anomalies on D-branes and I-branes result from the
chiral asymmetry of massless fermions on them.  
These fermions are in one-to-one correspondence with 
the ground states of the relevant open
string Ramond sectors.  In the case of N D-branes wrapping $M$, 
the 
relevant open strings start and end on identical but possibly 
distinct D-brane.  
Open string quantization\footnote
	{See the appendix for a discussion of the issue 
	of stability and supersymmetry of brane configurations.}
requires that the Ramond 
ground states be the sections of the spinor bundle lifted from 
$T(X) = T(M) + N(M)$, tensored with a vector bundle 
in the $(N, \bar N)$ representation (adjoint) of 
the gauge group $U(N)$ on the brane.  The latter
is dictated by the usual Chan-Paton factors.  
Because the adjoint 
representation is real, these fermions are CPT self-conjugate.  
We shall be interested in perturbative gauge anomalies, so
consider $\dim(M)$ to be even.  The GSO projection restricts the 
fermions to have a definite $SO(1,9)$ chirality.  
If $N(M) = \nil$, 
one is dealing with D9-branes.  The worldvolume theory  
is the super-Yang-Mills part of the type I string theory 
\cite {pol-RR}.  It is 
chiral and anomalous but its anomaly is cancelled by that of the  
gravitinos and the inflow from the close string sector 
via the Green-Schwarz 
mechanism \cite {gs}.  

When $N(M) \neq \nil$, the fermions have 
the quantum number $(+, +) \oplus 
(-, -)$ under the worldvolume Lorentz group $\Spin(1, p)$
and the spacetime Lorentz group restricted to $N(M)$: 
$\Spin (9-p)$.  The latter is now the global R symmetry of the 
worldvolume theory. 
If $N(M)$ is flat, left and right moving fermions as 
sensed by the worldvolume are treated equally and the 
theory is nonchiral.  However, when $N(M)$ has curvature, 
chiral asymmetry on the worldvolume is induced.  The 
point is that the worldvolume chiralities of the fermions 
are correlated with their representations  
under the global R symmetry.  Therefore a distinction 
arises between $(+, +)$ and $(-, -)$.  
The resulting perturbative anomaly can be
calculated by the family index theorem 
\cite {as, asz, sumi, agg, sz}.
For $\dim(M) = 4k+2$, the $(+, +)$ and $(-, -)$ fermions are
independent and separately Majorana.
The total anomaly associated with them is 
\beqar
	I_{D-brane} &=& 
	\frac {2 \pi} 2 \int_M \bigg( \ch[U(N)_{(N, \bar N)}] 
		\wg  \hat A[M] \nono 
	& & \wg \Big(\ch[S_{N(M)}^{+}] 
	- \ch[S_{N(M)}^{-}] \Big)
	  \bigg) ^ {(1)}.
\eeqar
Here $\ch[E]$ denotes the Chern
character of a vector bundle $E$.  $U(N)_{(N, \bar N)}$ 
denotes the vector bundle 
in the $(N, \bar N)$ 
representation of the structure group $U(N)$ 
associated with the N D-branes .
$S_{N(M)}^{ \pm}$ is the spin bundle lifted 
from $N(M)$ with $\pm$ chirality. $\hat A$ is the Dirac
genus.  
The factor of $\half$ in front reflects the reality
of the fermions.  
Since $U(N)$ is unitary, 
\beqar
	\ch[U(N)_{(N, \bar N)}] & = & \ch(F) 
		\wg \ch(-F^*) \nono
	& =  & \ch(F) \wg \ch(-F),
\eeqar
where 
\beq
	\ch(F) \equiv \exp (\frac F {2\pi}).
\eeq
$F$ is the properly normalized Hermitian field strength 
for the $U(N)$ 
connection on the D-brane in the \emph {fundamental} 
representation.  
Using  
\beq
	\ch [S_E^+] - \ch [S_E^-] 
	= \frac {e(E)} {\hat A(E)},
\eeq
which holds for any spin and orientable real vector bundle 
$E$, one can rewrite the anomaly as  
\beq
	\label {eq:d-brane-anomaly}
	I_{D-brane} = \frac {2 \pi} 2 
		\int_M \paren {\ch(F) \wg \ch (-F)
			\wg \hAtnd {} \wg e[N(M)]} ^ {(1)}.
\eeq
In the special case when $N(M)$ is null, $e[N(M)]$
as well as $A[N(M)]$ is 1.

	For $\dim(M) = 4k$, $(+, +)$ and $(-, -)$ are both 
complex and related by conjugation.  Anomaly can
be calculated by the contribution from either (but should not be
doubly  counted) as
\beqar	\label {eq:d-brane-anomaly-temp}
	I_{D-brane} &=& {2 \pi} 
	\int_M \Big( \ch(F) \wg \ch (-F) 
	\wg  \hat A(M) \wg 
	 \ch[S_{N(M)}^{+}] \Big)^{(1)} \nono
	&=& {2 \pi} 
	\int_M \bigg( \ch(F) \wg \ch (-F) 
	\wg  \hat A(M) \nono
	 &&\wg \half \Big(
	 \ch[S_{N(M)}^{+}] + \ch[S_{N(M)}^{-}] \nono
	 && + \ch[S_{N(M)}^{+}] 
	 	- \ch[S_{N(M)}^{-}] \Big)
	\bigg)^{(1)}.
\eeqar
Because $\ch [S_{N(M)}^{+}] +  \ch [S_{N(M)}^{-}]$ is a sum 
of Pontrjagin classes, it is made up of forms of ranks in
multiples of 4.  The same is true $\ch(F)\wg \ch(-F)$.  So only 
$\ch[S_{N(M)}^{+}] -  \ch[S_{N(M)}^{-}]$ can contribute 
in \eqr {d-brane-anomaly-temp} and
we obtain \eqr {d-brane-anomaly} again as the expression for
the anomaly.

	When two  D-branes intersect, additional massless
fermions arise from the open string sectors with two ends on
the two D-branes respectively.
Consider a
configuration with $N_1$ D-branes wrapping around $M_1$ and
$N_2$ around $M_2$.  In the sector with the string 
starting on $M_1$ and ending on $M_2$, 
the difference in the boundary conditions on the two ends of
the string modifies its zero point energy and shifts the
moding of some of its worldsheet operators 
\cite {pol-wit, bdl}.  The result is that the massless
fermions are a section of the chiral spinor bundle lifted
from 
\[T(M_1) \cap T(M_2) \oplus N(M_1) \cap N(M_2), \]
tensored with the $(N_1, \bar N_2)$ vector bundle due to their
Chan-Paton quantum numbers.  The anomaly can be 
calculated in the same
fashion as before:
\beqar	\label {eq:i-brane-anomaly}
	I_{I-brane} &=& 
	2 \pi \int_{M_{1 2}} 
		\bigg(\ch(F_1) \wg \ch(-F_2) 
		\wg \hAtni \nono
	  & &	\wg e [N(M_1) \cap N(M_2)]\bigg) ^ {(1)}.
\eeqar
Since 
$(N_1, \bar N_2)$ is complex, the fermions are not 
self-conjugate, and there is no factor of $\half$ in front.  
Note that \eqr {d-brane-anomaly}is precisely one half of 
the special case 
of \eqr {i-brane-anomaly} with 
$M_1 = M = M_2$.

	Using brane currents and \eqr {def-tau-M}, we can 
rewrite the anomalies \eqr {d-brane-anomaly} and \eqr
{i-brane-anomaly} in forms that will prove useful:
\beqar	\label {eq:d-brane-anomaly-preferred-X}
	I_{D-brane} & = & \pm \frac {2 \pi} 2 
		\int \tau_{M} 
	\wg \bigg( e[N(M)] \nono 
	& &\wg \ch(F)\wg \ch(-F) \wg\hAtnd{} \bigg)^{(1)}, \\
	\label {eq:i-brane-anomaly-preferred-X}
	I_{I-brane} 
	  & = & \pm 2 \pi \int \tau_{M_{1 2}} 
	  	\wg  \bigg( e[N(M_1) \cap N(M_2)] \nono 
	 & & \wg \ch(F_1) \wg \ch(-F_2) \wg \hAtni \bigg)^{(1)}.
\eeqar
Here we have left their signs  
undetermined because, being integrals of differential forms, 
they really depend on choices of orientation that are not yet 
fixed by any consideration so far.  
This ambiguity will soon be resolved by the requirement of 
factorizability.

In \cite {ghm}, the cases in which $M_{12}$ 
is the transversal intersection of 
$M_1$ and $M_2$, i.e. $N(M_1) \cap N(M_2) = \nil$, 
were considered.  Then the 
expression for I-brane anomaly \eqr {i-brane-anomaly} can be
further simplified as 
\beq 	\label {eq:i-brane-anomaly-preferred-X-nt}
	I_{I-brane} = 
	  \pm 2 \pi \int \tau_{M_1} \wg \tau_{M_2}  \nono
	   \wg \paren {\ch(F_1) \wg \ch(-F_2) \hAtni} ^ {(1)},
\eeq
where we have evaluated $e(\nil)$ to be $1$ but kept 
$\hat A([N(M_1) \cap N(M_2)]$ for future comparison.

	It is easy to check that 
\eqr {i-brane-anomaly-preferred-X-nt} 
is factorizable in the sense of 
\eqr {model-anomaly}, with 
\beq	\label {eq:soln-Y}
	Y_i = \ch (F_i) \wg \sqrt \hAtnd i
\eeq
and 
\beq	\label {eq:soln-tilde-Y} 
	\tilde Y_j = - (-1)^{\frac {dim(M_j) - q} 2} 
		\ch (-F_j) \sqrt {\hAtnd j}.
\eeq
Hence this anomaly can be cancelled by the 
inflow \eqr {gvary-s-12}.
The sign factor in \eqr {soln-tilde-Y} is determined by 
\eqr {def-tilde-Y}.  As promised before, this fixes the 
choice of orientation for the anomaly, and 
\eqr {i-brane-anomaly-preferred-X-nt} becomes 
\beqar 	\label {eq:i-brane-anomaly-preferred-a}
	I_{I-brane} &=& 
	  - \pi \int \tau_{M_1} \wg \tau_{M_2}  
	   \wg \bigg( \Big( (-1)^{\frac {dim(M_2) - q} 2}
	   	\ch(F_1) 
	   \wg \ch(-F_2) \nono 
	    && + \{1 \swap 2\} \Big) \wg \hAtni \bigg) ^ {(1)}.
\eeqar
After some manipulation one can show that the two terms in the 
integrand of \eqr {i-brane-anomaly-preferred-a} contribute 
equally, rather than cancelling each other, to the anomaly:
\beqar 	\label {eq:i-brane-anomaly-preferred-b}
	I_{I-brane} &=& 
	  - (-1)^{\frac {dim(M_2) - q} 2} 
	  2 \pi \int \tau_{M_1} \wg \tau_{M_2}  
	   \wg \bigg( \ch(F_1) \wg \ch(-F_2) \nono 
	   && \wg \hAtni \bigg) ^ {(1)}.
\eeqar

	\eqr {i-brane-anomaly-preferred-X-nt} 
is also trivially correct when $N(M_1)
\cap N(M_2)$ is nonempty but trivial, because the RHS' of both 
\eqr {i-brane-anomaly} and 
\eqr {i-brane-anomaly-preferred-X-nt} vanish.  However, 
\eqr {naive-intersection} would want one to believe that 
\eqr {i-brane-anomaly-preferred-X-nt} 
fails for a nontrivial $N(M_1) \cap N(M_2)$ because its 
RHS would seem to vanish, 
although the anomaly does not in general.  
There are similar difficulties for the D-brane anomaly 
\eqr {d-brane-anomaly}.  Consider D-branes with worldvolume 
$M$. For $N(M) = \nil$, the anomaly is that of Type I string theory 
and cancelled via the Green-Schwarz mechanism \cite {gs}.  
For $N(M) \neq \nil$, the 
closest thing would be \eqr {gvary-s} 
with $M_1 = M = M_2$.  However, 
$\tau_M \wg \tau_M$ naively vanishes.

\section {Topology to the Rescue}

It is clear from the earlier discussions that 
factorizability in the sense of \eqr {model-anomaly} 
is crucial for an anomaly to be cancelled 
via this inflow method.  However,
when the relevant normal bundle is 
nontrivial, it can be shown  that 
the \emph {integrand} of 
\eqr {i-brane-anomaly} is no longer factorizable because of 
the Euler class.  In other words, 
it is not factorizable 
unless $N(M_1) \cap N(M_2)$ is empty.  
The same can be said about the D-brane 
anomaly \eqr {d-brane-anomaly}.  
A related puzzle on 
the other side of the inflow mechanism has also been shown.  
The second equation in 
\eqr {naive-intersection} would imply 
vanishing inflow for $M_{1 2}$  
as long as $N(M_1) \cap N(M_2) \neq \O$, 
regardless of the twisting of the normal
bundle.  It could cancel no anomaly, factorized or not. 

	The origin of all these difficulties can be traced 
back to the properties of brane currents.  
Being a physical observable, $\tau_M$ must
be globally defined over $M$.  However, \eqr
{naive-current}
only makes sense within each coordinate patch, because
between patches the transversal coordinates are defined only
up to the transition functions for the normal bundle.
To it one must add additional terms, which vanish when 
$N(M)$ is trivial but turn 
$\tau_M$ into a globally defined form when $N(M)$ is not.  
Therefore
if such correction can be found, it must carry topological
information about $N(M)$, and from \eqr {naive-current} it
must have components with indices tangential to $M$. 
Mathematicians have found an elaborate construction for this
correction \cite {bt}.  By pulling $\tau_M$ back to $M$,
only parts from the correction can survive.  It is
remarkable that the result is cohomologically the Euler class 
$e[N(M)]$ of $N(M)$.

	Before proceeding further it is convenient to 
introduce some notations.  First observe that $\tau_M$ is
determined by $N(M)$, because it should be defined as the 
limit of nonsingular differential forms with shrinking 
compact supports in the neighborhood of $M$, which is 
approximated by the neighborhood of the zero section of
$N(M)$.  As such $\tau_M$ can be defined for any oriented
real orientable vector bundle $E$ by taking $M$ to be the zero section 
$E$.  To emphasize this we define\footnote 
{Actually for our purpose, knowledge of the cohomology class
of $\Phi (E)$ is sufficient. It is called the Thom class of
$E$.}
\beq	\label {def-Thom-form}
	\Phi [E] \equiv \tau_M
\eeq
for any vector bundle $\pi, E \to M$.  The important
property just mentioned can be written as
\beq	\label {eq:Thom-Euler}
	\tau_M \wg \tau_M = \tau_M \wg \Phi [N(M)] 
	= \tau_M \wg \Big[e[N(M)] \Big]
\eeq
where $[e]$ denotes some representative of the cohomology 
class of $e$.
Another useful property is
\cite {bt}:
\beq	\label {eq:thom-class-factorize}
	\Phi ( A \oplus B) = \Phi (A) \wg \Phi (B).
\eeq
This can be seen as Euler class also factorizes under
Whitney sum.  Now by \eqr {TX-factorize}, for the
I-brane worldvolume $M_{12} = M_1 \cap M_2$ we have 
\beqar	\label {eq:transversal-intersection}
	\tau_{M_1} \wg \tau_{M_2} & = &
	\Phi \brak {T(M_1) \cap N(M_2)
		\oplus N(M_1) \cap N(M_2) } \nono
	& & \wg  \Phi \brak {N(M_1) \cap T(M_2) 
		\oplus N(M_1) \cap N(M_2) } \nono 
	& = & \Phi \brak {T(M_1) \cap N(M_2) 
	\oplus N(M_1) \cap T(M_2) \oplus N(M_1)
	\cap N(M_2)} \nono
	& & \wg \Phi \brak {N(M_1) \cap N(M_2)} \nono
	& = & \tau_{M_{1 2}} \wg e [\brak {N(M_1) \cap N(M_2)}],
\eeqar
where in the last step we have used \eqr {thom-class-factorize} 
again along with \eqr {Thom-Euler}.
This is the correct replacement for the naive equation in 
\eqr {naive-intersection}.  Now returning to the I-brane anomaly 
\eqr {i-brane-anomaly-preferred-X}, one notes that as long as 
$\dim [T(M_1) \cap T(M_2)] + 2 
> \dim [N(M_1) \cap N(M_2)]$, 
one can use the freedom to add local counterterms to choose 
to make the Wess-Zumino descent on terms \emph {other than} 
the Euler form.  
The I-brane anomaly then becomes
\beqar	\label {eq:i-brane-anomaly-preferred-c}
	I_{I-brane} 
	  & = & \pm 2 \pi \int \tau_{M_{1 2}} 
	  	\wg  e\brak {N(M_1) \cap N(M_2)} \nono 
	 & & \wg \bigg( \ch(F_1) 
	 	\wg \ch(-F_2) \hAtni \bigg)^{(1)}.
\eeqar
By the same token, only the cohomology class of 
$e$ is important here.  
Substituting for \eqr {transversal-intersection}, one obtains 
again 
\eqr {i-brane-anomaly-preferred-X-nt} as the expression for 
anomaly.  But now it is clearly valid even when 
the normal bundle is nontrivial.  Furthermore, the
D-brane anomaly can also be written in this form with 
$M_1 = M_2 = M$, as long as $\dim [T(M)] + 2 
> \dim [N(M)]$.  When 
$\dim [T(M_1) \cap T(M_2)] + 2 
< \dim [N(M_1) \cap N(M_2)]$, both the anomaly and the inflow 
vanish.  The case of $\dim [T(M_1) \cap T(M_2)] + 2 
= \dim [N(M_1) \cap N(M_2)]$ is an intriguing one and we
will comment on it shortly.  
We have shown that except for that case, 
the inflow \eqr {gvary-s} not only does 
not vanish identically but cancels precisely the anomalies 
\eqr {i-brane-anomaly-preferred-X} and 
\eqr {d-brane-anomaly-preferred-X}. 

	There is a nice topological 
characterization of our results.  It has emerged 
that the anomaly, written as an
integral over the total spacetime, is always proportional to 
\beq	
	\tau_{M_1} \wg \tau_{M_2}.
\eeq
Its cohomology class is the Poincare dual of the
transversal intersection of $M_1$ and $M_2$.  Transversal
intersection, unlike geometric or set-theoretic intersection, 
has the property of stability: because there is
no common transverse direction, 
small perturbation can only move the intersection around 
but never make it disappear.  Consider now a 
\emph {nontransversal} intersection 
$M_{12} = M_1 \cap M_2$.  Because  
$N(M_1) \cap N(M_2) \neq \O$, 
a small perturbation in those directions would naively separate
them and lift the intersection altogether.  
This is the meaning of the second line in 
\eqr {naive-intersection}.  Such perturbation is given
by a global section of $N(M_1) \cap N(M_2)$.  However,
a global section of a sufficiently twisted vector bundle will 
necessarily have nonempty zero locus.  For 
$N(M_1) \cap N(M_2)$, this means that $M_1$ and $M_2$
cannot be completely separated.  Any small perturbation will
leave intact some submanifold of $M_{1 2}$, the zero locus
of the corresponding 
section of $N(M_1) \cap N(M_2)$, which is now stable.  
That is precisely the \emph {traversal} intersection 
of $M_1$ and $M_2$.  It can
be shown that the Poincare dual of the zero locus of an
orientable real vector bundle 
$E$ is none other than $e(E)$.  This gives
another derivation of \eqr {transversal-intersection}.  For
$M_1 = M = M_2$, the story is similar.  $e[N(M)]$ is the
Poincare dual of the zero locus of $N(M)$.  So 
$\tau_M \wg \tau_M$ measures the self-intersection of $M$.  
To recapitulate, D-brane and I-brane anomalies are
associated with transversal intersections, even when the 
pertinent geometric intersections are not transversal.  
In light of this, 
it seems worthwhile to introduce the notion of 
\nterm {transversal I-brane}, whose brane current is simply 
$\tau_{M_1} \wg \tau_{M_2}$.

	Now turning to the special case of 
\[ \dim [T(M_1) \cap T(M_2)] + 2 
	= \dim [N(M_1) \cap N(M_2)].	\]  
This implies that 
$\dim[T(M_1)] + \dim[T(M_2)] = 8$, or that the two D-branes make up  
an electro-magnetic dual pair.  An example would be 
a D-string intersecting with a D5-brane at 0 angle. 
For $M_1 = M = M_2$, the condition $\dim [T(M)] + 2 
= \dim [N(M)]$ means one is dealing with 
the self-dual D3-brane in IIB theory.
For these examples the anomaly 
\eqr {i-brane-anomaly-preferred-X} is finite 
but the inflow, even after 
taking into account the nontriviality of the normal bundles, 
still seems to vanish.  But one should not rush to conclude 
that anomaly does not cancel for them, because 
the intersection of electric and magnetic sources 
introduces an additional subtlety: 
the Chern-Simons action \eqr {c-s-action}
is no longer well defined.  A more powerful approach is 
needed but will not be pursued in the present work.  

\section	{Induced Brane Charges}

An important consequence of the inflow mechanism, besides 
lending support to
the consistency of various brane configurations, is that charges
for the bulk Ramond-Ramond fields are induced by the gauge fields
and gravitational curvatures as in \eqr {eom}.  Let $M$ be 
the worldvolume of some Dp-branes
with gauge field strength $F$.  Consider a
m-cycle $\gamma$ of $M$\footnote 
{In this section we count in complex unit the dimensions
of compactification manifolds $S$ if it is Calabi-Yau and in
real units those of other types as well as all
submanifolds of $S$.}.  Then
\beq	\label {eq:def-induced-charge}
	\Qind = \int_\gamma \ch(F) 
	  \wg \sqrt {\frac {\hat A[T(M)]}
	  		   {\hat A[N(M)]}}
\eeq 
gives the induced charge, in integral unit,  for the Ramond-Ramond 
$(p+1 - m)$-form potential.  From the viewpoint of the field
theory on the Dp-brane, the characteristic class on 
the RHS measures the topological charge of a
gravitational/Yang-Mills ``instanton''.  Let us
call it $Y$ as before.  Then \eqr {eom} shows that $\tau_M
\wg Y$ can be thought of as the brane current for a ``fat''
D$(p-m)$-brane bound to and spread out on the Dp-brane. 
When the instanton shrink to zero size, $Y$ also acquires
Dirac's $\delta$ singularity.  $\tau_M \wg Y$ is just like a
brane current.  One might well wonder if the instanton
can be lifted off the brane and become a physical D-brane in
its own right.  At least for Yang-Mills instantons there 
has been much evidence in support of this idea: field theory
instantons and branes are continuously connected by 
transitions between different branches of the moduli space 
of the I-brane field theory
\cite {Tsi,Tbwb,Tgfab,vafainst}.  Recently,
more complicated configurations involving gravitational
curvatures on the D-brane were used to study geometric
engineering and realizations of field theory dualities 
employing brane configurations \cite {bv,ho}.  In this
section we consider specific examples in which the twisting 
of the normal bundle modifies the induced charge.

	As discussed in the appendix, our analysis seems to 
apply, a posteriori, to nonsupersymmetric 
brane configurations as well. 
However, in most applications considered in the literature there 
are some supersymmetries left so as to have  
control over radiative corrections.  Therefore 
here we shall only consider Type II compactifications over
$d$-dimensional manifolds $S$ that preserve some 
supersymmetries.  A D-brane wraps around a $m$-dimensional 
submanifold $M$ of $S$
can preserve some of the supersymmetries of the 
compactification provided $M$ satisfies some conditions.  Such
a $M$ is called a supersymmetric cycle \cite {bbs}.  All 
supersymmetric cycles 
have been analyzed and classified in \cite {ooy,
bbmooy}.  We shall consider them one by one.  We shall also
only consider $S$' with irreducible holonomy because the
analysis for the other cases can be reduced to them.  
The forms in $\hat A(N)$ all have
ranks in multiples of $4$.  On the other hand, to have
nontrivial normal bundle, the D-brane must wrap a \emph
{proper} submanifold of $M$.  By counting dimensions and ranks, 
the contribution of $N(M)$ to $\Qind$ comes from the
rank $4$ component of  
\[	\frac 1 {\sqrt {\hat A(N(M))}}.	\]  
For convenience we shall group it 
together with the contribution from
$T(M)$ at the same rank, so the 
characteristic class we shall be computing is 
\beq	\label {eq:def-lambda}
	\lambda \equiv \ofe {p_1[N(M)] - p_1 [T(M)]}.
\eeq
Let the Chern roots of $T(M)$ be 
\beq
	\pm x_i, \come i = 1 \ldots \frac m 2.
\eeq
For the cases considered here $m$ is always even.  Let the
Chern roots of $N(M)$ be 
\beq
	\pm y_j, \come j=1 \dots \lrfloor {\frac {d-m} 2},
\eeq
with an additional 0 if $d-m$ is odd.  Then 
\eqr {def-lambda} can be written via the splitting principle
as 
\beq	\label {eq:lambda-x}
	- \oofe \paren {\sum_i x_i^2 - \sum_j y_j^2}.
\eeq
Of particular interest is whether $\lambda$ and hence
$\Qind$ can be expressed purely in terms of $x$'s, information
which is encoded in $T(M)$.

	The first nontrivial compactification is
$\Kt$.  However, for this case there cannot be any additional
contribution to the induced brane charge from a twisted
normal bundle, for dimensional reasons mentioned above.

	The next case is for $S$ to be a generic Calabi-Yau
$3$-fold.  According to \cite {ooy} a
supersymmetric cycle is either a Lagrangian submanifold 
(3-cycle) or a Kahler submanifold (2n-cycle) 
of the Calabi-Yau
3-fold.  For the reasons discussed above, only for Kahler
4-cycles does $N(M)$ make a contribution to $\Qind$.  The
holonomy of $T(M)$is $U(2)_T$ and that of $N(M)$ is
$U(1)_N$.  The Calabi-Yau condition requires 
\beq
	x_1 + x_2 + y = 0.
\eeq
The relevant charge is proportional to 
\beqar
	\lambda & = & \ofe {p_1[N(M)] - p_1 [T(M)]} \nono
	& = & \ofe {2 x_1 x_2} = \ofe {2 e(T(M))}.
\eeqar

	The remaining type of Calabi-Yau compactification is
over a generic Calabi-Yau 
4-fold.  It can have three types of supersymmetric
cycles: Lagrangian (4-cycle), Kahler
(2n-cycle), and Cayley  
(4-cycle).  A special
Lagrangian submanifold has the property that the holonomy of its normal
bundle is the same as that of its tangent bundle.  Therefore
the effect of $N(M)$ on the induced charge 
completely cancels whatever contribution from $T(M)$:  
$\lambda = 0$.

	Among the Kahler (2n)-cycles, $4$-cycles and
$6$ cycles will see contribution from $N(M)$.  The holonomy
group of $T(M)$ is $U(n)$.  The holonomy group of $N(M)$ is
$U(4-n)$.  The Calabi-Yau condition says that 
\beq
	\sum_i x_i + \sum_j y_j = 0.
\eeq
Using this we can calculate 
\beqar	\label {eq:lambda-kahler}
	\lambda 
	  & = & \oofe \paren 
	  	{2 \paren{\sum_{i_1 < i_2} x_{i_1} x_{i_2}
	  		- \sum_{j_1 < j_2} y_{j_1} y_{j_2}} }
	  	\nono
	&=& \ofe {2 c_2[T_+(M)] - 2 c_2[N_+(M)]}.
\eeqar
where $T_+(M)$ and $N_+(M)$ are the holomorphic tangent and
normal bundles of $M$ respectively, and $c_2$ denotes the second
Chern class.  For a Kahler $6$-cycle, $c_2[N(M)]$ is $0$, 
so \eqr {lambda-kahler} is entirely determined by information
encoded in $T(M)$.  This is not so for a Kahler $4$-cycle,
for which \eqr {lambda-kahler} reduces to 
\beq	\label {eq:lambda-kahler-4-cayley}
	\lambda_{4-\literal{cycle}} 
	  = \ofe {2 \Big( e[T(M)] - e[N(M)] \Big)}
\eeq
but cannot be expressed in terms of $x$ alone.

	Calabi-Yau 4-folds admit one more type of
supersymmetric cycles \cite {bbmooy}.  It is to date the
only known case where a single D-brane breaks the
supersymmetries of a type II compactification 
by $\frac 3 4$ instead of
$\half$.  They are known as Cayley submanifolds \cite {hl}.  
They are 4-dimensional 
and satisfy the conditions \cite {mclean, bbmooy}
\beq	\label {eq:cayley-cond1}
	x_1 + x_2 + y_1 + y_2 = 0,
\eeq
and
\beq	\label {eq:cayley-cond2}
	x_1 - x_2 = y_2 - y_1.
\eeq
These conditions are sufficiently restrictive 
to imply the vanishing of $\lambda$.

	There are two other cases of string
compactifications: $S$ may be a seven dimensional manifold
with $G(2)$ holonomy or an eight dimensional manifold with
$\Spin (7)$ holonomy \cite {sv, pt}.  A generic $\Spin (7)$
manifold supports only Cayley submanifolds as supersymmetric
cycles \cite {bbmooy}.  It is again 4-dimensional.  With 
a suitable choice of orientations, the curvature is   
subject to \eqr {cayley-cond1} but not \eqr {cayley-cond2}. 
Then \eqr {lambda-kahler-4-cayley} follows again \cite {bbmooy}.

	Finally we come to the case of $G(2)$ manifold.  It
admits two types of supersymmetric cycles \cite {bbmooy}. 
They are known as coassociative ($4$-cycle) and associative
($3$-cycle) submanifolds respectively.  Only for the
coassociative submanifold will $\Qind$ be affected by the
gravitational curvature.  With a suitable choice of orientations, 
they satisfy the condition 
\cite {mclean} that 
\beq
	x_1 + x_2 + y = 0.
\eeq
Hence 
\beq
	\lambda = \ofe {2 x_1 x_2} = \ofe {2 e(T(M))}.
\eeq

	The results in this section are summarized in the 
following table.
\begin {center}
\begin {tabular} {||l|l|c||}	\hline
Holonomy of S & Type of M & $\lambda$	\\ \hline
SU(3) & Kahler 4 & $2 e(M) / 48$	\\ \hline 
SU(4) & Special Lagrangian & 0 \\ \hline
SU(4) & Cayley	&  0 \\ \hline
SU(4) & Kahler 4 & $2 [e(M) - e(N)] / 48$ \\ \hline
SU(4) & Kahler 6 & $2 c_2 [T_+ (M)] / 48 $ \\ \hline
G(2) & Coassociative &	$2 e(M) / 48$	\\ \hline
Spin (7) & Cayley & $2 [e(M) - e(N)] / 48$ \\	\hline
\end {tabular}
\end {center}

\section* {Acknowledgment}
	We would like to thank K.~Bardakci, J.~de~Boer, 
O.~Ganor, M.~Green, J.~Harvey, A. Hashimoto, K.~Hori, 
S.~Kachru, A. Kato, I.~Klebanov, M.~Krogh, S.~Mandelstam, 
\moore, H.~Ooguri, E.~Sharpe, D. Sorokin, 
Y.~Oz, and B.~Zumino for useful discussions.  ZY would 
especially like to thank H.~Ooguri for encouragement 
throughout this work.  The research of YKEC was supported by
NSF-PHY9600258 and  DOEFG02-91ER40671.  
ZY was supported by DOEAC03-76SF00098.

\appendix
 
\section{Comments on Brane Stability and 
Supersymmetry}

It is appropriate to address the issue of  
stability of brane configurations 
and its relevance to the anomaly analysis\footnote 
	{We would like to thank K. Bardakci for useful 
	conversations regarding this issue.}.  
For a generic brane configuration, there are forces between
nonparallel branes.  If they do not cancel, this configuration is
not stable.  One can no more trust string
perturbation theory in an unstable brane configuration than
one can trust perturbative expansion around a false
vacuum in field theory.  Anomaly calculations 
is in some sense more robust than many other perturbative 
calculations, but 
one must still know the correct spectrum of massless fermions 
in \emph {some} true vacuum to correctly compute the anomaly.  
Of course this was the original motivation 
for t'Hooft's anomaly matching conditions.  
In the above, we have relied on
string perturbation when we obtained the massless
fermion contents and their quantum numbers.  When the 
brane configuration is unstable, there is no known reason to 
expect a priori that such analysis captures correctly 
the spectrum.

	On the other hand, supersymmetry 
is the only general condition under which 
the forces between branes cancel.  
If supersymmetry is completely 
broken in a brane configuration, the latter is 
\emph {generically} unstable.  
For N identical D-branes to
preserve some supersymmetry in a string compactification, they
must wrap around the supersymmetric cycles classified in
\cite {ooy, bbmooy}.  Between a pair of D-branes, the pattern
of supersymmetry breaking depends on their relative
arrangement.  For the 
case of intersection at right angle, some 
supersymmetries survive provided that 
\cite
{ghm}
\beq	\label {eq:cond-i-brane-susy}
	\dim [T(M_1) \cap N(M_2)] 
	  + \dim [N(M_1) \cap T(M_2)] = 0 \mod 4.
\eeq
The expression on the LHS of this equation is sometime denoted 
$nd + dn$ in the literature 
because it is the number of spacetime coordinates 
for which the boundary condition of the relevant open string
is Neuman on one end and Dirichlet on the other. When \eqr
{cond-i-brane-susy} is not satisfied, 
anomaly calculation based on perturbative
string theory does not have to be 
reliable.  For example, if $nd+dn = 2$,
it may be shown that the force between the two D-branes is 
attractive.  It is believed that in this case there exists
a stable nonmarginal bound state \cite {pol}.  
There seems a priori to be no reason to expect that 
the correct degrees of freedom 
of the bound state to be obtained from a 
perturbative string analysis  carried out  at the unstable 
configuration.  

	On the other hand, \eqr {cond-i-brane-susy} 
was not needed in the analysis carried out in this paper.  
In fact it follows through as long as 
\beq
	\dim [T(M_1) \cap T(M_2)] 
	  + \dim [N(M_1) \cap N(M_2)] = 0 \mod 2,
\eeq
a condition satisfied by any pair of D-branes that can 
coexist in the same string theory.
This seems to suggest that even for nonsupersymmetric 
brane configurations, at least 
the massless fermion contents might be captured correctly.

\myref {
\bibitem {gs}	\green, \schwarz, 
	\plb 149,84,117.
\bibitem {ch} \callan, \harvey, \npb 250,85,427.
\bibitem {bh} J. Blum, \harvey, \npb416,94,119,hep-th/9310035.
\bibitem {ghm}  \green, \harvey, \moore, \cqg 14,97,47, 
		hep-th/9605033.
\bibitem {bsv2}	\bershadsky, \sadov, \vafa, 
		\npb463,96,420, hep-th/9511222.
\bibitem {ooy} \ooguri, Y. Oz, \zy, 
		\npb477,96,407, hep-th/9606112.
\bibitem {bbmooy} \beckerd, \morrison, 
		\ooguri, Y. Oz, Z. Yin, \npb 480,96,225,
		hep-th/9608116.
\bibitem {bsv1}	\bershadsky, \sadov, \vafa, 
		\npb463,96,398, hep-th/9510225.
\bibitem {ov1}	\ooguri, \vafa, \npb463,96,55, hep-th/9511164.
\bibitem {bv}	\bershadsky, A. Johansen, 
		T. Pantev, \sadov, 
		\vafa, 
		\npb448,95,166, hep-th/9612052.
\bibitem {ov2}	\ooguri, \vafa, hep-th/9702180. 
\bibitem {vz}	\vafa, \zweibach, hep-th/9701015.
\bibitem {ho}	\hori, \oz, hep-th/9702173. 
\bibitem {top1}	M.~Blau, G.~Thompson,
	\npb492,97,545, hep-th/9612143.
\bibitem {top2} L. Baulieu, A. Losev, N. Nekrasov, 
		hep-th/9707174.
\bibitem {witea} \witten, hep-th/9610234. 
\bibitem {bcr} L. Bonora, C. S. Chu, M. Rinaldi, hep-th/9710063.
\bibitem {Tsi}  \witten, \npb460,96,541, hep-th/9511030.
\bibitem {Tbwb} \douglas, hep-th/9512077. 
\bibitem {Tgfab} \douglas, hep-th/9604198. 
\bibitem {gh3}	P. Griffiths and J. Harris, 
	{\sl Principles of Algebraic Geometry}, Chap. 3, 
	Wiley-Interscience, 
	New York 1978.
\bibitem {wz} \wess, \zumino, \plb37,71,95.
\bibitem {zlec} \zumino, Lectures given at Les Houches 
	Summer School on Theoretical Physics, 1983.
\bibitem	{pol-RR} \polchinski, 
		\prl75,95,4724, hep-th/9510017.
\bibitem {dght} S. Deser, A. Gomberoff, M. Henneaux, C. Teitelboim,
		\plb400,97,80, hep-th/9702184.
\bibitem {pol} \polchinski, Lectures given at TASI Summer 
	School on Fields, Strings and Duality, 1996, 
	hep-th/9611040.
\bibitem {ss} \schwarz, \sen, \npb411,94,35, hep-th/9304154.
\bibitem {sorokin} G. Dall'Agata, K. Lechner, 
	D. Sorokin, hep-th/9707044.
\bibitem {wy} T. Y. Wu, C. N. Yang, \prd12,75,3845.
\bibitem {as} \atiyah, \singer, 
	\pnas81,84,2597.
\bibitem {asz} \alvarez, \singer, \zumino, \cmp96,84,409.
\bibitem {sumi} T.~Sumitani, \jpa17,84,L811.
\bibitem {agg} \agaume, \ginsparg, \annp161,85,423, 
	Erratum \ibid171,86,233.
\bibitem {sz} J.~Manes, R.~Stora, \zumino, \cmp102,85,157.
\bibitem {pol-wit} \polchinski, \witten, \npb460,96,525,
	hep-th/9510169.
\bibitem {bdl} M. Berkooz, \douglas, R. G. Leigh, \npb480,96,265, 
	hep-th/9606139.
\bibitem {bt}	R.~Bott, L.~W.~Tu, {\sl Differential Forms in 
		Algebraic Topology}, Chap. 2, Springer-Verlag, 
		New York, 1982.
\bibitem {vafainst}	\vafa, \npb463,96,35, hep-th/9512078.
\bibitem {bbs}	\beckerd, \strominger, 
		\npb456,95,130, hep-th/9507158.
\bibitem {hl}	R. Harvey, B. Lawson, \am148,82,47.
\bibitem {mclean}	R. C. McLean, 
	http://www.math.duke.edu/preprints/96-01.ps.
\bibitem {sv}	S. Shatashvili, \vafa, 
		\sm1,95,347, hep-th/9407025.
\bibitem {pt} G. Papadopoulos, P. K. Townsend, 
	\plb357,95,300, hep-th/9506150
}
\end {document}